# Direct Calculation of Modal Contributions to Thermal Conductivity via Green-Kubo Modal Analysis: Crystalline and Amorphous Silicon

Wei Lv (吕玮)[1], Asegun Henry[*,1,2]

**Affiliations:**

[1]George W. Woodruff School of Mechanical Engineering, Georgia Institute of Technology, Atlanta, GA 30332, USA.

[2]School of Materials Science and Engineering, Georgia Institute of Technology, Atlanta, GA, 30332, USA.

[*]Corresponding author: ase@gatech.edu

Abstract

In this letter we derive a new method for direct calculation of the modal contributions to thermal conductivity, which is termed Green-Kubo modal analysis (GKMA). The GKMA method combines the lattice dynamics formalism with the Green-Kubo formula for thermal conductivity, such that the thermal conductivity becomes a direct summation of modal contributions, where one need not define the phonon velocity. As a result the GKMA method can be applied to any material/group of atoms where the atoms vibrate around stable equilibrium positions, which includes not only crystalline line compounds, but also random alloys, amorphous materials and even molecules. By using molecular dynamics simulations to obtain the time history of each mode's contribution to the heat current, one naturally includes anharmonicity to full order and can obtain insight into the interactions between different modes through the cross-correlations. As an example, we applied the GMKA method to crystalline and amorphous silicon and because the modal contributions at each frequency result from the analysis, one can then apply a quantum correction to the mode heat capacity to predict the thermal conductivity at essentially any temperature. The temperature dependent thermal conductivity for amorphous silicon, computed using quantum corrected GKMA results, shows the best agreement with experiment to date and serves as a validation of the formalism. The GKMA method provides new insight into the nature of phonon transport, as it casts the problem in terms of mode-mode correlation instead of scattering, and provides a general unified formalism that can be used to understand phonon-phonon interactions in essentially any class of materials or structures where the atoms vibrate around stable equilibrium sites.

Main Text

The phonon gas model (PGM) originated from the behaviors observed and rationalized in homogenous crystalline solids and it has exhibited remarkable success in describing the behavior of a wide variety of solids, microstructures, nanostructures and molecules [1–4]. In homogenous crystalline solids (e.g., line compounds), where there is both compositional and structural periodicity, the dynamical matrix determined from lattice dynamics (LD) is homogenous with symmetric blocks and as a result, all of the eigen solutions correspond to plane wave modulated vibrations (e.g., propagating modes). Plane wave modulated vibrations exhibit a well-defined phase and often group velocity, because each wave has a clearly defined wavelength, and there is a clear dispersion relation. These eigen solutions/normal modes of vibration then transport heat as they propagate, with a clearly defined velocity, which is consistent with the PGM based physical picture of their transport. In the PGM each mode carries energy $h\upsilon$ with velocity $d\omega/dk$ and mean free path (MFP) $\Lambda$, which is the product of the velocity and the time between energy exchanges between modes (e.g., scattering events). This physical picture works well for homogenous crystalline materials and the thermal conductivity of virtually any solid compound and their corresponding nanostructures can now be computed from first principles [2,3,5–9]. Given its success, it has become the primary lens with which phonon transport is viewed.

The problem with ubiquitous usage of the PGM is that for systems that lack periodicity or homogeneity, the eigen solutions/normal modes do not in general correspond to plane wave modulated vibrations with a clearly identifiable wavelengths, and therefore one cannot define the phonon dispersion or velocity. The lack of a clearly defined velocity is critical, because the PGM hinges on the velocity being defined in order to properly describe a mode's contribution to thermal transport. Thus, for systems that lack periodicity, such as amorphous materials, random alloys or molecules, using the PGM to describe their phonon transport is inconsistent with the atomic level vibrations. This issue is critical, because these classes of materials represent a major fraction of the materials used in various applications that involve heat transfer.

One class of materials that have proved difficult to explain even with effective MFP based arguments is amorphous materials. Several existing theories [10–13] have worked towards resolving this long-standing issue, but they are all based on the PGM and are therefore based on a set of modes that are inconsistent with the modes of vibration in amorphous materials, which are largely non-propagating. To our knowledge the only exception has been the seminal work of Allen and Feldman (A-F), who made a major step forward by conducting LD calculations on supercells of amorphous Si [14]. In doing so, they developed a model for thermal conductivity that was, for the first time, based on the actual modes of vibration in an amorphous material, which resolved the major issue with previous work. A-F used the Kubo formula for thermal conductivity and calculated the contributions of different modes assuming the interactions between atoms were harmonic [14,15]. In this sense, it is important to note that they did consider anharmonicity to the extent that it is required for a temperature gradient to develop, corresponding to finite thermal conductivity. However, beyond this assumption, in their mathematical treatment of the atomic interactions, they treated them as purely harmonic.

Michalski [16], on the other hand, argued that anharmonicity is important and a more recent study showed that the A-F method is not accurate for many amorphous materials other than a-Si [17]. Recently He *et al.* [18] and Jason *et al.* [19] used normal mode analysis (NMA) to include anharmonic effects on the contributions of propagons to the thermal conductivity of a-Si. Since propagons only account for ~ 3% of the total number of modes, their contributions to thermal conductivity remains controversial, [20] even though their contributions may be large on a per mode basis.

Despite these concerns, to date, the A-F model has offered one of the best explanations for the thermal conductivity of amorphous silicon [14,15].

Towards the goal of incorporating temperature dependent anharmonicity to full order, we present such a formalism termed the Green-Kubo mode analysis (GKMA) method, as it is a combination of Green-Kubo and the normal mode analysis method. Following lattice dynamics formulism, we first examine the meaning of the reverse transformation from normal mode coordinates back to individual atom coordinates, where the normal mode amplitudes are calculated from,

$$X(n) = \sum_j \sqrt{m_j}\, \mathbf{p}_j^*(n) \cdot (\mathbf{x}_j - \mathbf{x}_{j0}) \tag{1}$$

and

$$\dot{X}(n) = \sum_j \sqrt{m_j}\, \mathbf{p}_j^*(n) \cdot \mathbf{v}_j \tag{2}$$

the displacement and velocity of each atom $\mathbf{x}$ and $\dot{\mathbf{x}}$ can be obtained from the normal mode coordinates via,

$$\mathbf{x}_j = \frac{1}{m_j^{1/2}} \sum_n \mathbf{p}_j(n) X(n) \tag{3}$$

$$\dot{\mathbf{x}}_j = \frac{1}{m_j^{1/2}} \sum_n \mathbf{p}_j(n) \dot{X}(n) \tag{4}$$

In Eqs. (1-4) $n$ denotes the mode (e.g., the $n^{th}$ solution to the equations of motion), $m_j$ is the mass of the $j^{th}$ atom, and $\mathbf{p}_j(n)$ is the polarization vector which gives the relative magnitude and direction of motion for atom j in mode $n$. Equations (1-4) are not new and are well established within the context of the LD formalism [21].

The critical enabling insight offered herein is the physical meaning associated with Eqs. (3) and (4). Here, we postulated that Eqs. (3) and (4) essentially state that at every instant, each atom's position and velocity are composed of their respective contributions from the different collective vibrations in the system. Thus, every atom's position and velocity are dictated by the respective magnitudes of each normal mode's amplitude $X_j(n)$ and its time derivative $\dot{X}_j(n)$. By thinking of each atom's position and velocity as being composed of an exact sum of modal contributions at every instant, it was then postulated that if an individual mode's contribution to the displacement or velocity of an atom is used in an expression for the calculation of any other property that depends on that atom's position and/or velocity, one would subsequently obtain that mode's contribution to that property. For example, one could calculate each mode's contribution to the temperature of the system as discussed in the supplementary materials. Similarly, towards the calculation of thermal conductivity, the modal contributions to the velocity of each atom can be substituted into the heat flux operator derived by Hardy [22], to obtain each mode's contribution to the volume averaged heat flux at each time step in a EMD simulation,

$$\mathbf{Q} = \sum_n^{3N} \mathbf{Q}(n) = \sum_n^{3N} \frac{1}{V} \sum_i \left[ E_i \mathbf{v}_i(n) + \sum_j (-\nabla_{\mathbf{r}_i} \Phi_j \cdot \mathbf{v}_i(n)) \mathbf{r}_{ij} \right] \tag{5}$$

where $\mathbf{v}_i(n) = \frac{1}{m_j^{1/2}} \mathbf{p}_j(n) \dot{X}(n)$, $E_i$ is the sum of potential and kinetic energy of atom i, V is the volume of the supercell, $\Phi_j$ is the potential energy of atom j, and $\mathbf{r}_{ij}$ is the distance between atom i and atom j.

We can then take this expression and substitute it directly into the Green-Kubo expression for thermal conductivity, to obtain an equation that expresses the thermal conductivity as a direct summation over individual mode contributions,

$$\kappa(n) = \frac{V}{k_B T^2} \int_0^\infty \langle \mathbf{Q}(n,t) \cdot \mathbf{Q}(0) \rangle dt \tag{6}$$

Furthermore, one can also substitute the summation of modal contributions to the heat flux in both places of the heat flux autocorrelation to obtain the thermal conductivity as a double summation over individual mode-mode heat flux correlation functions,

$$\kappa = \frac{V}{k_B T^2} \int_0^\infty \left\langle \sum_n \mathbf{Q}(n,t) \cdot \sum_{n'} \mathbf{Q}(n',0) \right\rangle dt = \frac{V}{k_B T^2} \sum_{n,n'} \int_0^\infty \langle \mathbf{Q}(n,t) \cdot \mathbf{Q}(n',0) \rangle dt \tag{7}$$

Equations (6) and (7) are the primary results termed the GKMA method, which now allows us to obtain each mode's contribution to the total thermal conductivity, and we are guaranteed by Eq. (4) that the summation in Eqs (6) and (7) will exactly recover the total GK thermal conductivity. Also, Eq. (7) allows one to examine how the correlation between pairs of modes contributes to thermal conductivity. The predominant viewpoint of phonon-phonon interactions is based on the PGM and is thought of in the context of scattering. However, in the GK formalism thermal conductivity arises from correlation [23], which is a fundamentally different way of viewing the physics. One would definitely expect that there is a strong connection between the two viewpoints, especially in the context of crystalline materials. However, it is important to appreciate that the GK representation of each mode's contribution to thermal conductivity through correlation is in and of it complete and does not require validation or correspondence to the prevailing PGM paradigm based on scattering. From this perspective, one can plot the magnitude of the individual terms in the double summation in Eq. (7) as elements of a 2D matrix, whereby the magnitude of the element represents the extent of correlation between modes. We postulate that a strong mode-mode correlation suggests the two modes somehow interact strongly, frequently or for long periods of time and possibly in collaboration with other modes.

Another useful feature of the GKMA method is that one can reduce the computational cost by combining any desired group of modes together to calculate the summed contribution of their combined heat flux at one time, which mitigates the need to separately output and store the heat flux associated with each mode. The details are discussed in the supplementary materials.

Aside from these benefits the main power of the GKMA method (e.g., Eqs. (6) and (7)) is that it now allows for calculation of the eigen mode contributions to thermal conductivity directly without the need to define a phonon velocity. This is a critical issue for situations where the PGM becomes questionable, such as for disordered materials. With the GKMA method, however, it is now possible to calculate the thermal conductivity contributions of individual modes for any arbitrary collection of atoms, as long as they vibrate around stable equilibrium sites.

The key question is whether or not the fundamental postulate by which the GKMA method is based on is in fact correct; specifically, is it true that the individual terms of the sum in Eq. (6) correspond to the actual modal contributions to thermal conductivity? To answer this question we compared this interpretation of the GKMA results with other well established methods for crystalline silicon, since it has been studied extensively [2,3,7]. Figure 1(a) shows the thermal conductivity accumulation for silicon using Eq. (6), which is in agreement with previous work [7] using the environment dependent interatomic potential (EDIP) [24] as well as first principles calculations density functional theory (DFT) [3]. As both EDIP and Tersoff are a short ranged empirical potentials [25] the agreement between them is better than with DFT, which includes longer ranged interactions and is more accurate. The simulation details are given in the supplementary information. Figure 1(b) shows the contributions of different polarizations to the total thermal conductivity, which is also consistent with the other two methods. This correspondence serves as validation that Eqs. (6) and (7) do in fact correspond to the thermal conductivity contributions associated with a given mode.

Based on this initial validation, we applied the GKMA method to a-Si, which is a system that cannot be well described by previous methods. The IPR indicates the extent to which a mode is localized and does not involve all of the atoms in the system as a widespread collective vibration. Propagons and diffusons are delocalized and therefore have small IPR. Locons, on the other hand, are localized vibrations and therefore exhibit high IPR, which in Fig. 2(a) manifests at the higher frequencies.

We then applied the GKMA to analyze the modal contributions to thermal conductivity in a-Si for all modes where propagons, diffusons and locons are all treated the same way via Eq. (7). The normalized thermal conductivity accumulation function vs. phonon frequency for a-Si is shown in Fig. 2 (c). For comparison, the accumulation from the A-F method is also shown, which was calculated using the implementation in the General Utility Lattice Program (GULP) [26].

In Fig. 2(c), the GKMA result, which includes all degrees of anharmonicity, predicts a similar trend as the A-F result at room temperature, which did not include anharmonicity. At first this might seem to suggest that anharmonicity is not important. However, examination of the 2D cross-correlation terms (e.g., Eq. (7)) shown in Fig. 3, indicates that there is significant correlation between modes with different frequencies. The harmonic A-F model, on the other hand, would only predict diagonal terms, and all off diagonal terms should be zero, due to the presence of the delta function $\delta(\omega_i - \omega_j)$ in their derivation [14]. Figure 3(a) shows the mode-mode cross correlations. The diagonal terms are the largest, but they only account for ~ 70% of the total thermal conductivity at room temperature. Therefore cross-correlations, which arise due to anharmonicity, are responsible for approximately 30% of the thermal conductivity. What is also remarkable about the result in Fig. 3 is the fact that there is a distinct change in the magnitude of the correlations around 16 THz. This transition coincides with the transition to localized modes (e.g., locons – see Fig. 3). Here it is important to note that no information regarding the nature of the modes (e.g., propagon, diffuson, or locon) was used to generate Fig. 3. Every mode in Fig. 3 was treated identically and no filtering was used to highlight the feature at 16 THz. Instead, a natural feature in the mode-mode correlations arises at the frequency where the mode character switches from spatially delocalized to localized. In Figs. 3(b) and (c) we have filtered out the auto-correlations (cross-correlations only) and cross-correlations (auto-correlations only) respectively to make the features more clear. Figure 3(c) shows that the locons do not have strong auto-correlations and the accumulation in Fig. 2(c) is consistent with previous assertions that locons exhibit a negligibly small contribution to thermal conductivity. One result of the A-F analysis is that the correlations between modes (e.g., interactions) should be most significant when the frequency of two modes is close. However, using the GKMA approach, Fig. 3(b) suggests that interactions between modes of many different frequencies can all interact, as there is no obvious increase as one approaches the diagonal. It is interesting to note, nonetheless, that despite these differences, both GKMA and the A-F model yield a similar normalized accumulation plot (see Fig. 2(c)). The same mapping of mode-mode correlations for c-Si is presented in the supplementary materials.

In comparing the magnitudes of thermal conductivity values produced by GKMA and the A-F method, the A-F result underestimated the total thermal conductivity of amorphous silicon by 30% at 300K. Here the most relevant experimental result is taken to be the measurements of Cahill et al. [27], which included the minimum hydrogen concentration (1%). Since the GKMA method allows us to determine the thermal conductivity accumulation with respect to phonon frequency, we can apply a quantum correction to the classical MD GKMA results at different temperatures and compare to the experimental data at all temperatures. The underlying assumption in doing so is that the only aspect of the problem MD cannot capture is the quantum effect on the specific heat. Applying a quantum correction to low temperature MD data, assumes that somehow the mode-mode interactions are still modeled correctly, despite the fact that the mode amplitudes/occupations are incorrect. Here, we test this assumption by applying a quantum correction to the GKMA results, which only imparts a correction to the specific heat component of each mode's thermal conductivity contribution via the ratio of the quantum expression for the volumetric specific heat [1] and the classical expression for the volumetric specific heat. See Eq. (S5) in supplementary materials.

Figure 4(c) shows a comparison of the quantum corrected thermal conductivity using GKMA as compared to experiments, which shows best agreement of all work to date [14,27,29]. The GKMA results in Fig. 4(c) were generated by a linear interpolation of the normalized thermal conductivity accumulation functions at 100K, 300K, and 800K. The un-normalized accumulations at each temperature are shown in Fig. 4(a) and (b), which indicates that a significant shift in the contributions occurs at lower temperatures. The fact that the non-quantum corrected MD result increases at lower temperatures shows that in order to obtain the correct temperature dependence, the frequency dependence must be obtained from GKMA first, so that the quantum correction can correctly scale each mode's contribution. At lower temperatures, the quantum correction nullifies the higher frequency contributions. However, it is still important to correctly calculate seemingly over predicted values with MD for the lower frequencies, which when quantum corrected exhibit excellent agreement with the experimental data [27].

In this letter we proposed a new method for calculating the modal contributions to thermal conductivity termed the GKMA method. We validated the fundamental postulate that the GKMA method is based on, by calculating the thermal conductivity of crystalline silicon, which showed good agreement with previous methods. We then applied the GKMA method to a-Si and obtained excellent agreement with experimental data, better than any previous model for a-Si. The major advantage of the GKMA method is that it can be applied systems where phonon velocities are not well defined and application of the PGM becomes questionable.

**Acknowledgments**

This research was supported Intel grant AGMT DTD 1-15-13 and computational resources were provided by the Partnership for an Advanced Computing Environment (PACE) at the Georgia Institute of Technology and National Science Foundation supported XSEDE resources (Stampede) under grant numbers DMR130105 and TG-PHY130049.

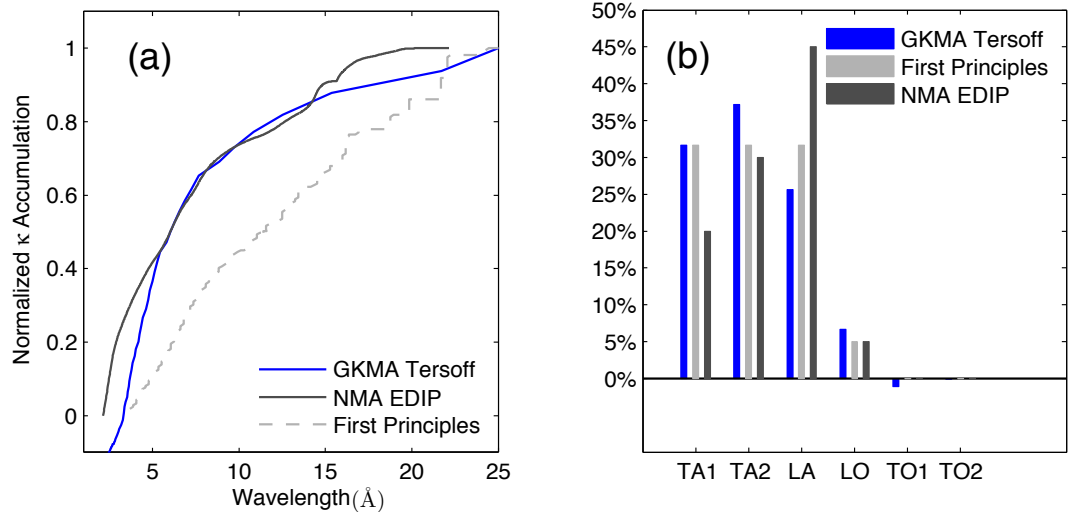

FIG. 1. (a) Thermal conductivity accumulation of c-Si with wavelength. (b) Comparison of each phonon polarization's contribution of c-Si.

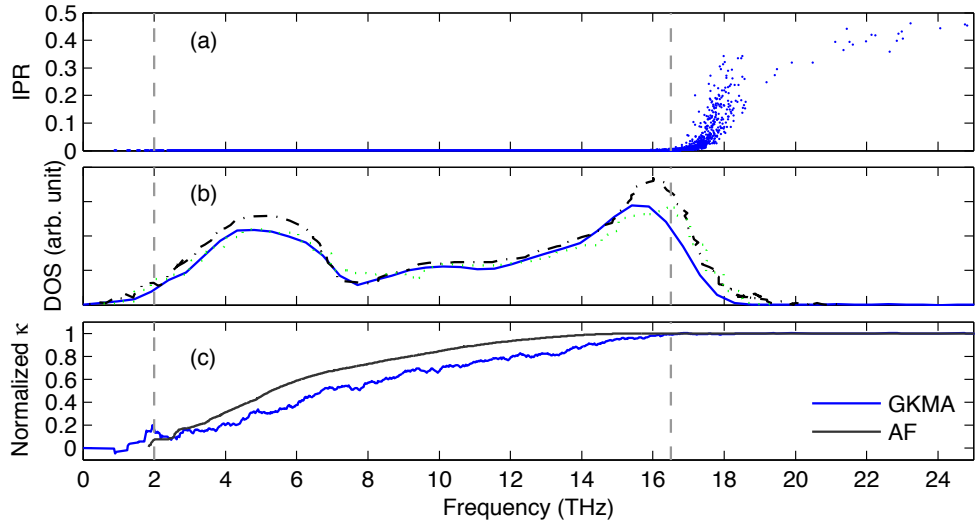

FIG. 2. (a) Inverse participation ratio of modes in a-Si; (b) Phonon density of states, black dash-dot curve is from [15] and green dotted line is from [17]; (c) Normalized thermal conductivity accumulation vs. mode frequency for a-Si using GKMA and A-F theory at 300K. The dotted gray lines are estimated cut-off between propagons and diffusons & between diffusons and locons.

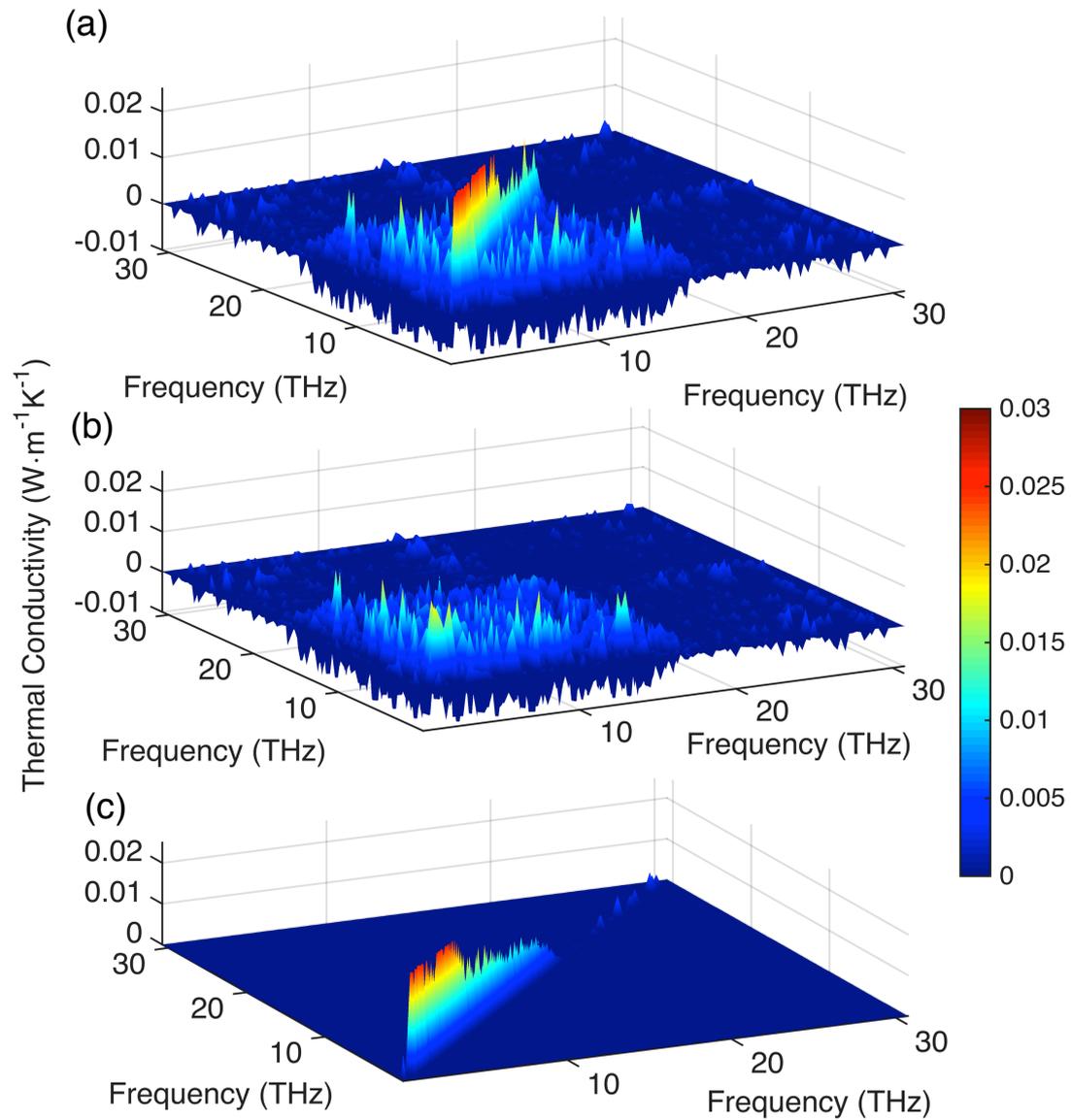

FIG. 3. (a) Thermal conductivity (TC) contributions from mode-mode correlations of amorphous silicon; (b) TC contributions from just mode-mode cross-correlations; (c) TC contributions from only mode-mode auto-correlations.

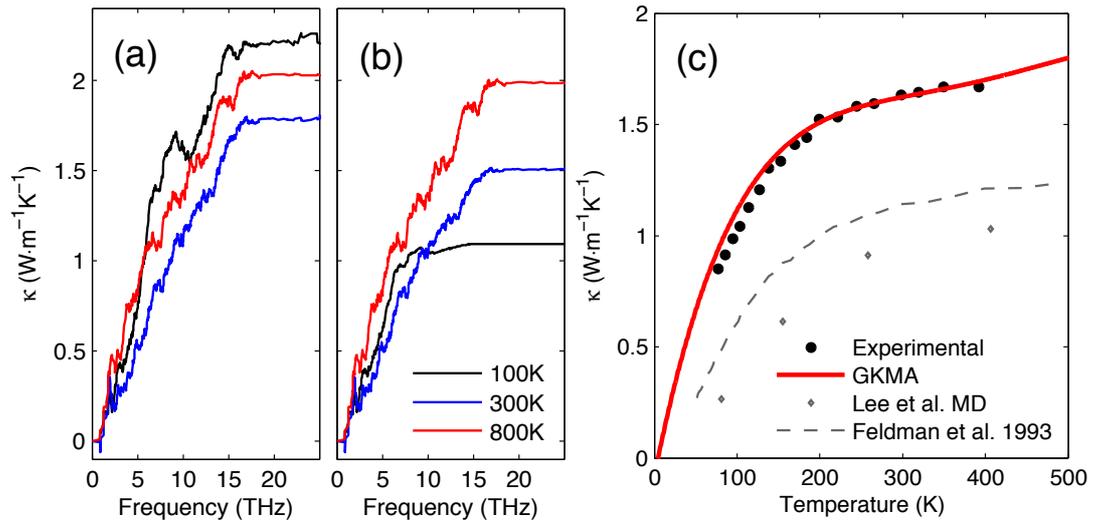

FIG. 4. Thermal conductivity accumulation of a-Si at 100K, 300K and 800K without (a) and with (b) quantum correction; (c) Thermal conductivity vs. Temperature for a-Si comparing with experiments [27] and simulation results from other methods [14,29].